\documentclass[12pt,thmsa,a4paper,notitlepage]{article}
\usepackage{sw20aip}

 \input tcilatex
\QQQ{Language}{
American English
}

\begin{document}

\title{The role of $S_{11}$ resonance in $\pi N$ Scattering}
\author{$^1$Yohanes Surya, $^2$Kam Chan Hin, $^2$T.C Au Yeung, \and $^2$W.Z.
Shangguan, $^3$Wirawan Purwanto, $^4$Eddy Yusuf , \and $^5$Darmadi Kusno, $%
^2 $Oki\ Gunawan \vspace{1cm} \\
{\small $^1$Universitas Pelita Harapan, Indonesia}\\
{\small $^2$Nanyang Technological University, Singapore}\\
{\small $^3$ Physics Dept. College of William and Mary\ }\\
{\small $^4$Physics Dept University of Indonesia}\\
{\small $^5$Physics Dept. Florida State University}}
\date{}
\maketitle

\begin{abstract}
We analyze Pion Nucleon Scattering up to $700$ MeV using a simple,
relativistic, unitary model\cite{1}. The kernel of the integral equation
includes nucleon, roper, delta, $D_{13}$ as well as $S_{11}$ poles with
their corresponding crossed pole terms approximated by contact interactions.
The $s$- and $p$- wave phase shifts are calculated from the model and shown
to agree very well with the values derived from $\pi $N scattering data \cite
{2}. All parameters which involve $S_{11}$ are presented.
\end{abstract}

\section{Introduction}

Even though pion-nucleon scattering has been extensively studied for many
years, there still remains a number of interesting problems to be explored,
especially with the construction of powerful new facilities such as TJNAF
(Thomas Jefferson National Lab) in Virginia, USA. In this brief paper we
analyze pion nucleon scattering up to 700 MeV laboratory kinetic energy of
pion using a simple, covariant and unitary model of $\pi N$ scattering\cite
{1}.

In this work, the $\pi N$ scattering amplitude is obtained as a solution of
a relativistic wave equation in which the pion and eta is restricted to
their mass shell in all intermediate states. The rationale for this approach
has been discussed by Gross and Surya\cite{1}. The kernel of the
relativistic integral equation includes undressed delta ($\Delta ),$roper$%
(N^{*}),D_{13}$ and $S_{11}$ poles in addition to the undressed nucleon $(N)$
pole. The kernel also includes contributions derived from crossed $N,\Delta
,N*,D_{13},$and $S_{11}$ diagrams, as well as from $\sigma $- and $\rho $%
-like exchange terms. To keep the model simple the crossed terms are
approximated by contact interactions as described in reference [1].

The $\pi NN$ coupling is taken to be the superposition of both pseudoscalar
and pseudovector coupling and the $\pi NS_{11}$ coupling is taken as a
superposition of both scalar and vector coupling.\ However for simplicity we
take only scalar coupling for the $\eta NS_{11}$ coupling (the results
obtained also show this tendency). The Feynman rule for the $\pi NN$ and $%
\pi NS_{11}$vertex are given as follows,

\[
\pi NN\quad \text{vertex}:\quad \Gamma _{0N}=g_{\pi NN}\left( \lambda +\frac{%
1-\lambda }{2m}{\gamma }^\mu {k}_\mu \right) \gamma ^5\tau _i\quad 
\]

\[
\pi NS_{11}\quad \text{vertex}:\quad \Gamma _{0S}=i\left( g_1+\frac{g_2}m{%
\gamma }^\mu {k}_\mu \right) \tau _i
\]
\[
\eta NS_{11}\quad \text{vertex}:\quad \Gamma _{0\eta }=ig_\eta \tau _0
\]
$k$ is the pion momentum, $\lambda $ is a mixing parameter that is
determined by the requirement that the nucleon mass be unshifted by the
interaction \cite{1}, $g_1,$ $g_2$ and $g_\eta $ are the coupling constants
adjusted to fit data and $m$ is the nucleon mass. The $\pi NN$ coupling  $%
g_{\pi NN}$ is chosen to be equal to 13.5 the same as in the reference [1].

\section{The model}

The model has been described in detail in reference [1] except for the $%
S_{11}$ resonance. Figure (1) shows additional kernels that one needed to
describe the $S_{11}$ resonance. Consistent with the previous model, the
crossed diagram is approximated as a contact interaction.

Figure (2) shows the self energy of $S_{11}.$ $M_c$ is the infinite sum of
iterated the contact diagrams. The $M_c^{\pi \pi },M_c^{\pi \eta },M_c^{\eta
\pi }\quad $and$\quad M_c^{\eta \eta }$ are treated as a coupled
channel.\quad \quad \quad \quad \quad \quad \quad \quad \quad \quad \quad
\quad \quad \quad \quad \quad \quad \quad \quad

\FRAME{ftbpF}{4.9485in}{0.8484in}{0pt}{}{}{pis1.eps}{\special{language
"Scientific Word";type "GRAPHIC";maintain-aspect-ratio TRUE;display
"USEDEF";valid_file "F";width 4.9485in;height 0.8484in;depth 0pt;cropleft
"0";croptop "0.9949";cropright "0.9995";cropbottom "0";filename
'C:/SW20/GRAPHICS/PIS1.EPS';file-properties "XNPEU";}}\quad \FRAME{ftbpF}{%
3.9946in}{2.994in}{0pt}{}{}{pis2.eps}{\special{language "Scientific
Word";type "GRAPHIC";maintain-aspect-ratio TRUE;display "USEDEF";valid_file
"F";width 3.9946in;height 2.994in;depth 0pt;cropleft "0";croptop
"1";cropright "1";cropbottom "0";filename
'C:/SW20/GRAPHICS/PIS2.EPS';file-properties "XNPEU";}}\quad \quad \quad
\quad \quad \quad \quad \quad \quad \quad \quad \quad \quad

\section{Results}

Due to the contact interactions approximation, the model allows us to fit
the spin $\frac 12$and spin $\frac 32$ phase shifts separately. Our fit to $%
S_{11},S_{31},P_{11}$ and $P_{31}$ phase shifts are given in Figures 3 and
4. The fit is good. The $S_{11}$ phase shifts are particularly interesting.
Figure 5 shows how the total is built up from individual contributions, the
curves in the figures show the result when the kernel $(i)$ includes only
the direct nucleon pole term and the contact term derived from crossed
nucleon exchange $plus$ the combined $\sigma $- and $\rho $- like contact
terms (the dot line), $(ii)$ the terms in ($i$) $plus$ the additional $\rho $%
-like $\pi \pi NN$ contact term (the dot-dashed line), $(iii)$ the terms in (%
$ii$) plus roper driving terms (dashed line) and finally $(iv)$ the total
result, which includes the terms in ($iii$) plus $S_{11}$ driving terms
(solid line).

From our calculation the contribution from $S_{11}$ (including $\eta N\quad $%
channel) plays an important role to pion nucleon interaction at energies
above 600 MeV as shown in Figure 5.

Table 1 shows  parameters that are used to fit the data\ (column 2). The
table includes effective masses $(m)$ and widths$\quad (\Gamma )$ of the
resonances (column 3) . In the table, $\Lambda \quad $represents the baryon
cut off mass, $g\prime $ is the inelastic coupling in the roper channel. We
found the $S_{11}$ width is smaller compared to Particle data group ($\sim $%
100 to 250 MeV)\cite{3}. This maybe is related to our smaller value of g$%
_{\eta NS_{11}}=1.19\;$as compared to effective models results \cite{3} or
QCD sum rules using interpolating field results \cite{4}which is around 2.

\vspace{.5cm}

$
\begin{tabular}[t]{lll}
parameter & value & effective value \\ 
$\Lambda $ & 1205.8 &  \\ 
$\Lambda (N^{*})$ & 1961.3 &  \\ 
$g_{\pi NN^{*}}$ & 6.924 &  \\ 
$m_{N^{*}}$ & 1458.0\quad & 1463.5 \\ 
$\Gamma _{N^{*}}$ &  & 244.3 \\ 
$C_\rho $ & 0.911 &  \\ 
$g\prime \quad $ & 0.793 &  \\ 
$g_2$ & 1.910 &  \\ 
$g_1$ & $1.222$ &  \\ 
$\Lambda _{s11}$ & 2666.3 &  \\ 
$m_{s11}$ & 1540.6 & 1567.0 \\ 
$\Gamma _{s11}$ &  & $57.4$ \\ 
$g_{\eta NS_{11}}$ & $1.19$ & 
\end{tabular}
$

\section{Conclusions}

We have successfully extended the relativistic, simple and unitary model of
pion nucleon scattering \cite{1}to analyze the pion nucleon scattering up to
700 MeV which includes the $S_{11}$ resonance. We find that the model works
well despite of our meson on-shell approximation in all intermediate states.
We also find that the inclusion of $\eta N$ channel tends to give the
correct high energy behavior. The coupling of pion and eta to the nucleon
and $S_{11}$ is smaller than expected. This would be clarified by some quark
model calculations.

\section{Acknowledgment}

YS would like to thank School of Electrical \& Electronic Engineering \
Nanyang Technological University for their support and for kind hospitality
during the visit.\newpage\ 

\FRAME{ftbpFU}{310.125pt}{239.875pt}{0pt}{\Qcb{$S_{11}$ phase-shifts}}{}{%
Figure }{\special{language "Scientific Word";type
"GRAPHIC";maintain-aspect-ratio TRUE;display "USEDEF";valid_file "T";width
310.125pt;height 239.875pt;depth 0pt;cropleft "0";croptop "1";cropright
"1";cropbottom "0";tempfilename 'C:/SW20/TEMP/FJTQKJME.bmp';}}\FRAME{ftbpFU}{%
335.5pt}{259.3125pt}{0pt}{\Qcb{P - wave phase shifts}}{}{Figure }{%
\special{language "Scientific Word";type "GRAPHIC";maintain-aspect-ratio
TRUE;display "USEDEF";valid_file "T";width 335.5pt;height 259.3125pt;depth
0pt;cropleft "0";croptop "1";cropright "1";cropbottom "0";tempfilename
'C:/SW20/TEMP/FJTK0KXI.bmp';tempfile-properties "XP";}}\FRAME{ftbpFU}{335.5pt%
}{259.3125pt}{0pt}{\Qcb{S - wave phase shifts}}{}{Figure }{\special{language
"Scientific Word";type "GRAPHIC";maintain-aspect-ratio TRUE;display
"USEDEF";valid_file "T";width 335.5pt;height 259.3125pt;depth 0pt;cropleft
"0";croptop "1";cropright "1";cropbottom "0";tempfilename
'C:/SW20/TEMP/FJTK0K3P.bmp';tempfile-properties "XP";}}

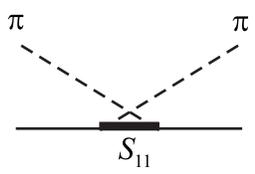
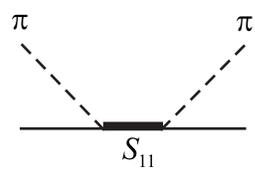
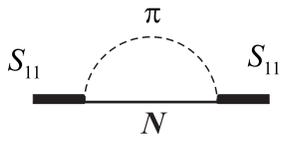
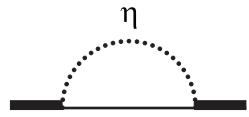
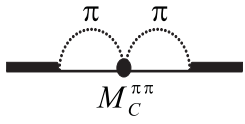
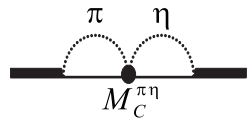
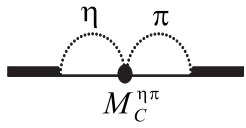
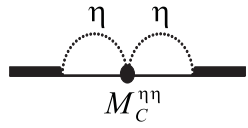
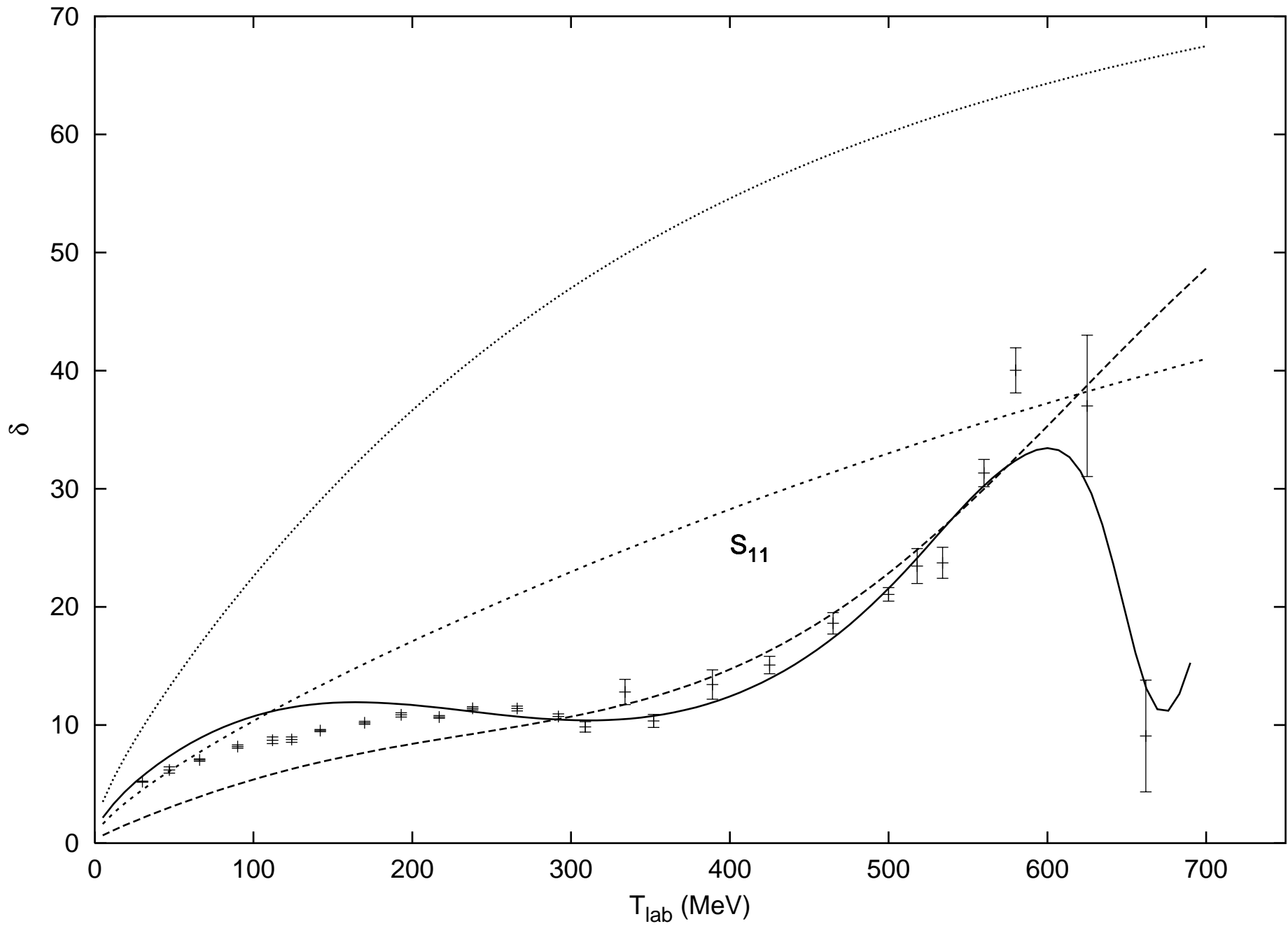
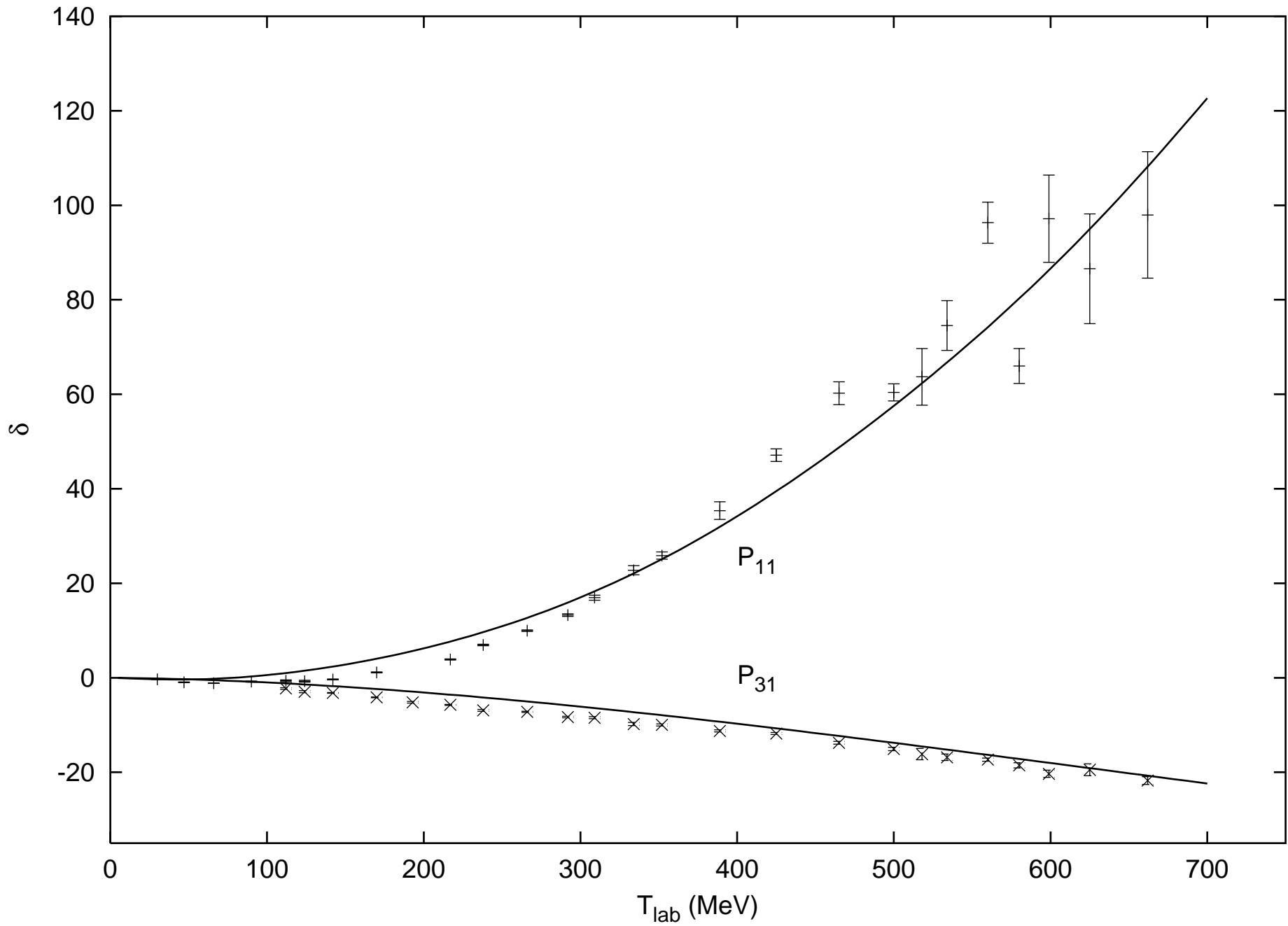
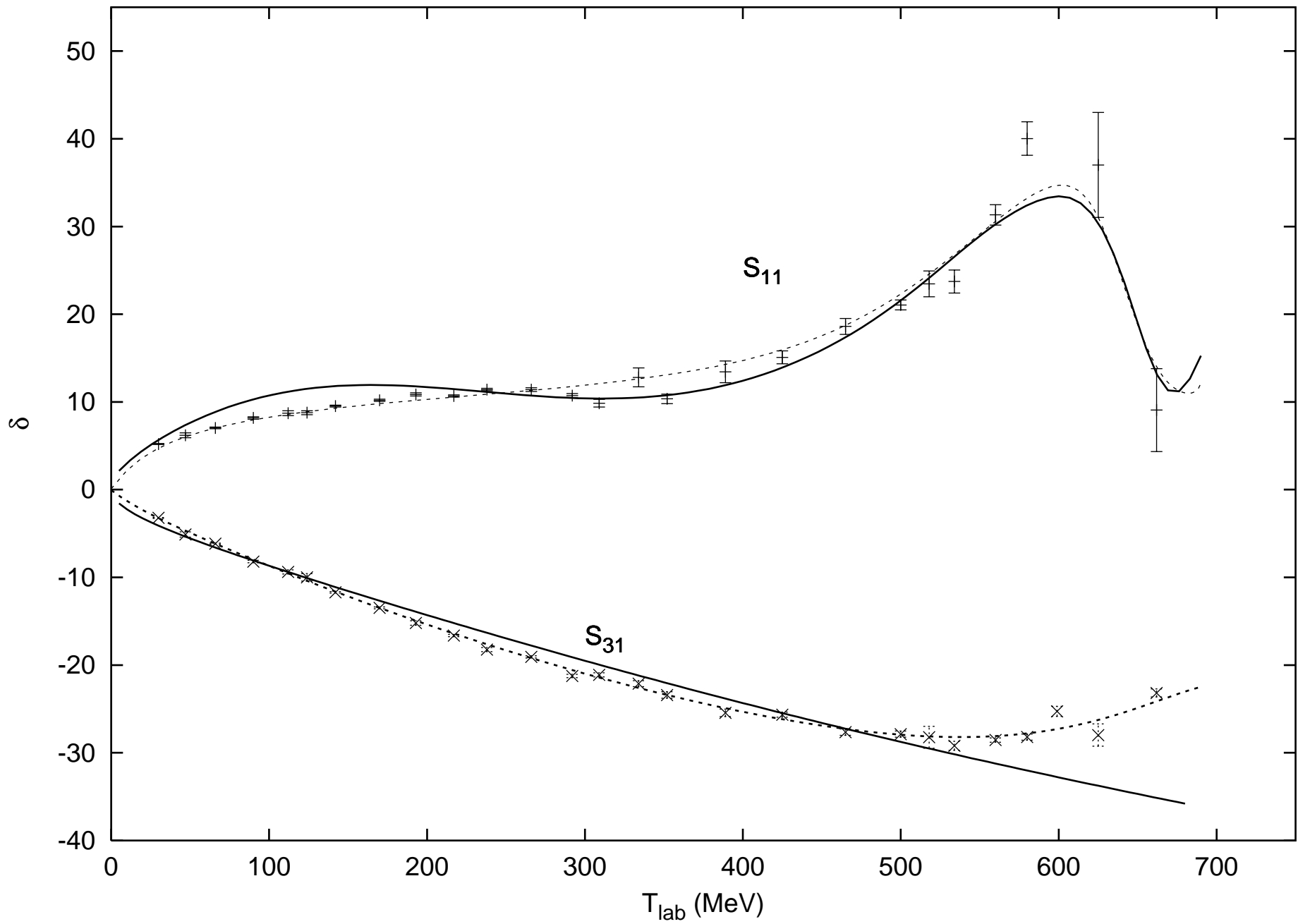
\end{document}